# 基于 AI 的电力电子变换器开路故障诊断研究综述

刘闯[1]，寇磊[1]，蔡国伟[1]，赵梓含[2]，张哲[3]

（1. 东北电力大学 电气工程学院，吉林省 吉林市 132012；
2. 国网新源丰满培训中心，吉林省 吉林市 132108；
3. 丹麦科技大学 电气工程学院，丹麦 灵比 2800）

Review for AI-based Open-Circuit Faults Diagnosis Methods in Power Electronics Converters

LIU Chuang[1], KOU Lei[1], CAI Guowei[1], ZHAO Zihan[2], ZHANG Zhe[3]

(1. School of Electrical Engineering, Northeast Electric Power University, Jilin 132012, Jilin Province, China;
2. State Grid Xinyuan Fengman Training Center, Jilin 132108, Jilin Province, China;
3. Department of Electrical Engineering, Technical University of Denmark, Lyngby 2800, Denmark)

**ABSTRACT:** Power electronics converters have been widely used in aerospace system, DC transmission, distributed energy, smart grid and so forth, and the reliability of power electronics converters has been a hotspot in academia and industry. It is of great significance to carry out power electronics converters open-circuit faults monitoring and intelligent fault diagnosis to avoid secondary faults, reduce time and cost of operation and maintenance, and improve the reliability of power electronics system. Firstly, the faults features of power electronic converters are analyzed and summarized. Secondly, some AI-based fault diagnosis methods and application examples in power electronics converters are reviewed, and a fault diagnosis method based on the combination of random forests and transient fault features is proposed for three-phase power electronics converters. Finally, the future research challenges and directions of AI-based fault diagnosis methods are pointed out.

**摘要：**电力电子变换器已被广泛应用于航空航天系统、直流输电、分布式能源及智能电网等领域，其可靠性问题成为学术界和工业界的研究热点。开展对电力电子变换器开路故障监测与智能诊断方法研究，对避免二次故障、降低运维时间和成本和提高电力电子系统可靠性具有重要意义。文章首先对电力电子变换器故障特征进行分析与总结，然后对特定基于人工智能技术的电力电子变换器开路故障诊断方法及应用实例综述，并提出一种基于随机森林与瞬时故障特征相结合的故障诊断方法用于三相电力电子变换器故障诊断。最后提出基于 AI 的电力电子变换器开路故障诊断方法所面临的挑战，并展望该领域未来的研究方向。

**关键词：**人工智能；电力电子变换器；开路故障；故障诊断；故障定位；IGBT；数据驱动

**DOI：10.13335/j.1000-3673.pst.2019.2427**

## 0 引言

随着新能源发电及其电能输送、工业电动机驱动、电动汽车、轨道交通等新兴产业的迅速崛起，电力电子技术已经成为能源变换的关键技术[1-3]。目前电力电子变换器已被广泛应用于电力系统、航空航天系统、不间断电源(uninterrupted power supply，UPS)、无功补偿、直流输电以及分布式能源等领域，其可靠性问题成为学术界和工业界的研究热点[4-6]。电力电子系统的可靠性变得至关重要，因此研究电力变换器中的功率器件故障诊断方法并对其进行及时处理，对提高电力电子系统的可靠性具有重要意义。

尽管有各种手段可以提高电力电子系统的可靠性，但是电力电子变换装置的运行环境日益复杂，特别是受极高温或低温、高湿度、强腐蚀、高压强等极端运行环境的影响，故障是不可避免的[7-8]。通常大型风电机组多运行在恶劣的环境下，机组故障停运维护费用高、维修时间长。2007年荷兰滨海埃赫蒙德海上风电场由于电气系统故障(变流器、控制系统等)导致的停运率较高，造成巨大的经济损失[9]。2011年徐矿综合利用发电有限公司中存在由于三相电流变送器、有功功率变送器烧坏的问题，使得 UPS 电源的电压、电流、及频率不正常；此外还存在整流器工作电流增大 40%左右，而保护装置没有检测到的情况，使得整流器长时间带故障运行而烧坏，从而导致 UPS 系统故障[10]。HXD3 机车主变流柜的主电路分别由 PWM 逆变器、四象限整流器模块等构成，2016 年某机务段发生 18 起 IGBT 炸裂故障，其中机车主变流柜内逆变侧和整流侧的 IGBT 发生爆炸事故较多[11]。因此开展对电力电子

基金项目：国家科技部重点研发专项(2017YFB0903300)。
Project Supported by National Key R&D Program of China (2017YFB0903300).

变换器状态监测和故障诊断的研究，及时发现故障并进行容错处理或者切换备用设备，能够避免停机影响生产的同时为维修人员提供设备故障信息。

电力电子变换装置主要由功率半导体器件组成，而功率半导体器件是电力电子变换装置中最脆弱的环节[12]。根据调查可知 IGBT 已经成为应用最广泛的功率半导体器件之一[13]。本文以 IGBT 故障为例进行分析，通常 IGBT 故障主要形式有短路故障和开路故障，其中短路故障具有很强的破坏性，基于软件算法来实现 IGBT 短路故障诊断保护的方法是不合理的，也是难以实现的，一般采用标准的硬件短路保护电路来完成[14-15]；而 IGBT 开路故障短时间内不会引发严重的过电流或者过电压现象，可以持续一段时间并且不会触发系统的保护，因此短时间内不会引起系统立即停机[16]。根据[17]可知由于电网电流不对称引起风机大规模跳机最短在几秒钟内，最长在几分钟内，但在恶劣的工况或者长时间带故障运行情况下，会导致 IGBT 开关器件发热量和损耗加大，开路故障会引起系统中其他元器件的二次故障，甚至引发灾难性的故障，导致较高的维修费用[18]。电力电子变换装置的电路数学模型总是不够精确，而且很多电路的数学模型非常复杂，在故障条件下就更难以建立精确可靠的故障数学模型。因此针对基于 AI 的电力电子变换器开路故障诊断方法的研究得到广泛关注。

基于人工智能的故障诊断方法通常不需要建立所研究对象的精确数学模型，按照有无监督学习可以分为监督学习、半监督学习、无监督学习，其中监督学习方法主要有：人工神经网络法(artificial neural network，ANN)[19-20]、支持向量机(support vector machine，SVM)[21-22]、随机森林(random forests, RFs)[23-25]、K 最近邻算法(k-nearest neighbor，KNN)[26]等；半监督学习方法有：生成式方法[27]、基于分歧的方法[28]等；无监督学习方法主要有自组织神经网络(self organizing map，SOM)[29]、聚类算法[30]、主成分分析法(principal component analysis，PCA)[31][32]等；还有一些基于深度学习、强化学习、迁移学习算法的故障诊断算法等，其中基于深度学习的方法主要有卷积神经网络(convolutional neural network，CNN)[33-34]、深度置信网络(deep belief network，DBN)[35]、堆叠自编码器(stacked auto encoders，SAE)[36-37]等方法；基于强化学习的方法主要有 Q 学习(Q-learning)[38]等方法；基于迁移学习的方法主要有 TrAdaBoost[39]等方法。此外人工智能技术在电力电子系统的自动化设计、状态监测、控制等领域也得到广泛应用[40-45]。本文首先对电力电子变换器中故障特征及部分保护方法进行简单介绍，然后对一些基于人工智能的电力电子变换器开路故障诊断方法及应用实例进行介绍，并提出一种基于随机森林与瞬时故障特征相结合的故障诊断方法，并将该方法应用于三相电力电子变换器的开路故障诊断。最后指出未来面临的挑战及人工智能技术在电力电子变换器开路故障诊断领域可能的研究方向。

# 1 电力电子变换器主要故障

电力电子变换器故障主要由功率半导体器件故障引起，功率半导体器件故障主要有短路故障和开路故障 2 种形式[46]。电力电子变换器被广泛应用到智能电网、航空航天、高速铁路、以及新能源发电等诸多领域(图 1)，因此研究电力电子变换器中 IGBT 故障成为当前国内外关注的焦点问题之一[47-50]。

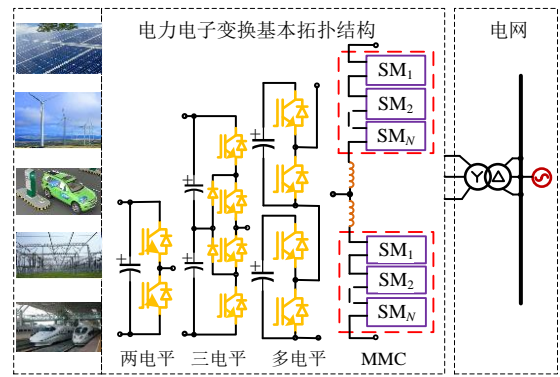

**图 1 电力电子变换器的一些应用领域**
**Fig. 1 Some application fields of power electronics converters**

## 1.1 短路故障

短路故障主要是由雪崩、过热、过压击穿以及错误驱动信号等原因导致，IGBT 短路故障发生后的主要特点是在非常短的时间内容易引起巨大的冲击电流，破坏性强，容易烧坏电力电子装置的其它元器件[51]。IGBT 短路故障保护通常由硬件来实现，包括退饱和检测法、阻容分压网络检测法、电感检测法、吸收钳位电路保护法、集电极电流检测法、饱和压降精确测量法和 IGBT 慢关断保护法等[52]。文献[53]提出一种通过分析门极电压来实现 IGBT 过流检测的方法(图 2)，当检测到短路故障后立即封锁信号，IGBT 将表现出开路故障的特征。文献[54-55]利用快速熔断器(fuse)热容量小，在故障电流未达到预期的短路电流时即可被熔断的特性，在逆变器桥臂中串联 2 个快速熔断器将短路故障转化为

开路故障(图3)，从而降低其危害。

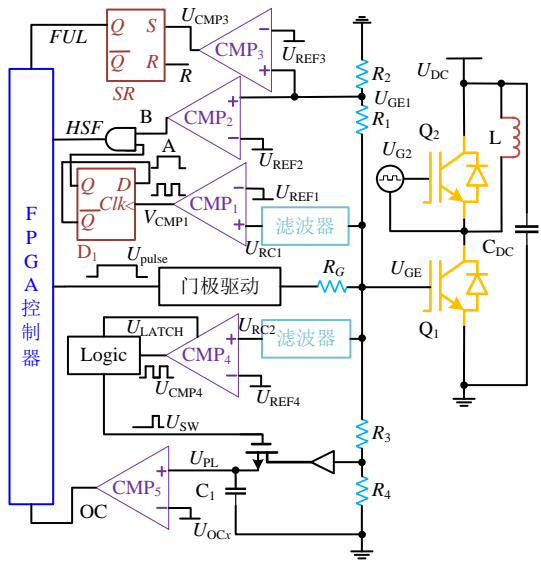

**图 2 过流保护电路**
**Fig. 2 Over-current protection circuit**

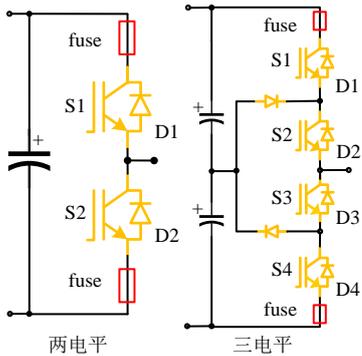

**图 3 快速熔断器短路故障隔离技术**
**Fig. 3 Short-circuit fault isolation technology with fast fuses**

### 1.2 开路故障

IGBT 开路故障的主要原因有器件破裂、绑定线断裂或焊接脱落、接线不良、驱动断线或电路失效等，或者任由 IGBT 短路也会导致 IGBT 烧毁而形成开路故障[56-57]。根据文献[57-60]可知，通常 IGBT 开路故障后(图 3 中 S1)，与其反并联的二极管(图 3 中 D1)仍然能够正常工作，系统不会立即崩溃，但性能会降低，会导致电流电压谐波含量增加，降低供电质量，但 IGBT 开路故障可能长时间不被发现，从而导致其他设备的二次损坏或者灾难性的故障发生[61-62]。

#### 1.2.1 基于解析模型的开路故障诊断方法

基于解析模型的电力电子开路故障诊断方法通常是利用电力电子变换系统的数学模型，通过监测电力电子变换器故障时系统的电压、电流等信息与估计值进行比较得到偏差来实现 IGBT 开路故障诊断(图 4)。文献[63]对电机驱动系统中多相电压源逆变器中功率管开路故障进行研究，提出基于极电压误差标准化的故障诊断方法，该方法主要依据不同故障状态下逆变器输出极电压与正常工作状态下的偏差来确定开路故障的位置，并采用阈值比较器来消除测量误差及噪声的影响。文献[64]提出一种简单的用于三相 PWM 逆变器中 IGBT 开路故障诊断方法，将电流畸变与阈值进行比较，结合重新计算得到的电流矢量旋转角来实现故障位置的诊断。文献[65]提出一种基于线电压误差的永磁直驱风电系统变流器开路故障诊断方法，通过比较正常状态下和故障状态下线电压得到偏差，将线电压误差值与设定的阈值进行比较来实现故障定位，并利用电压幅值和时间宽度双重标准来提高诊断方法的可靠性。基于解析模型的故障诊断方法通常对阈值的设定要求比较高，否则会使得诊断精度降低。

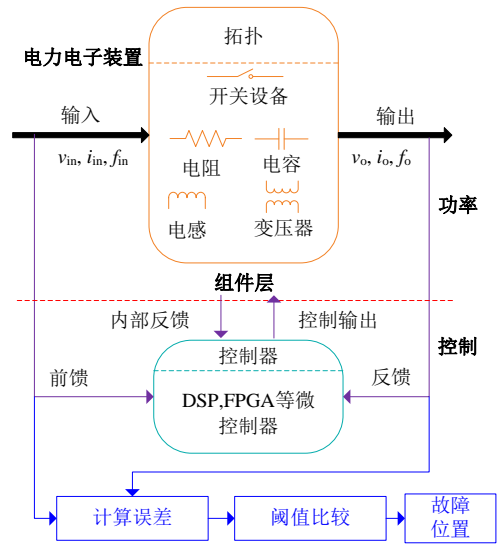

**图 4 基于解析模型的开路故障诊断方法**
**Fig. 4 Open-circuit fault diagnosis method based on analytical model**

#### 1.2.2 基于人工智能的开路故障诊断方法

电力电子变换装置主要由功率半导体器件组成，而包含功率半导体器件的系统并不是线性系统。因此，线性控制理论和理想的半导体开关模型在电力电子变换器的开路故障方法的研究中并不完全适用，从而限制基于故障数学模型的电力电子变换器开路故障诊断方法的推广应用。目前针对电力电子变换装置故障建模的研究极少，无法提供电路的故障机理模型[33]，而基于数据驱动的人工智能技术具有模拟任何连续非线性函数的能力和从故障样本自适应学习的能力，其主要利用特定人工智能算法实现对故障数据与故障状态之间的映射，通过对历史数据进行离线学习和挖掘得到成熟的故障诊断分类器(图 5)，然后再利用训练成熟的故障

诊断分类器来实现对电力电子变换系统的在线故障诊断[66-67]。

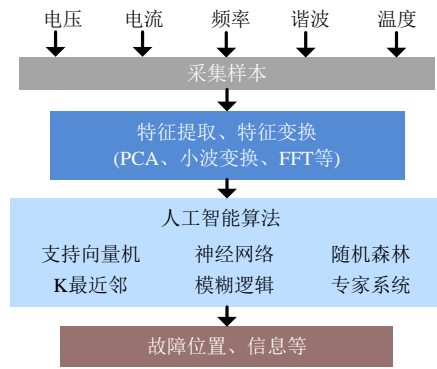

**图 5 基于人工智能的开路故障诊断方法**

**Fig. 5 Open-circuit fault diagnosis method based on Artificial Intelligence**

文献[22]针对三相 PWM 逆变器中 IGBT 开路故障特征进行研究，通过监测三相电流计算得到电流矢量轨迹，并提取均值、方差等特征来训练 SVM 故障诊断分类器，从而实现对 IGBT 开路故障的定位与诊断。

文献[68]提出一种基于人工神经网络的用于串联混合动力汽车电源交直流变换器故障诊断的方法，首先通过分析证明直流母线电流是采集的最佳数据，直流母线电流携带有与电机电流、电机电压和直流母线电压相对应的有价值信息，再利用快速傅里叶变换(fast Fourier transform，FFT)对原始信号进行特征提取，最后将数据用于训练神经网络得到故障诊断分类器，从而实现开路故障的定位。但是 FFT 变换不可避免会产生频谱能量泄露与栅栏效应，影响谐波分析的准确性。傅里叶变换作为一种全局性的变换，其具有一定的局限性。而小波变换是空间(时间)和频率的局部化分析，它通过伸缩平移运算对信号逐步多尺度细化，能够实现自适应时频信号分析的要求，能够克服傅里叶变换窗口不能随频率变化的缺点。文献[69]对电磁法三电平变换器中开关开路故障进行研究，利用小波包分析法提取变换器的输出电压信号，然后利用核主成分分析法对特征进行降维处理得到训练数据，再利用概率神经网络进行训练得到故障诊断分类器，从而准确地诊断出逆变器中开路故障的位置。

文献[24]中提出一种基于数据驱动的残差分析方法用于实现风力发电机组故障诊断与隔离，利用长短期记忆网络(long short-term memory，LSTM)来处理所有状态变量的连续原始数据来产生残差，然后再利用残差数据训练随机森林，从而实现对故障的诊断和隔离，提高风机系统的稳定性，避免严重事故的发生，降低运维成本及维修成本。

文献[70]提出一种基于混合核支持张量机(mixed kernel support tensor machine，MKSTM)的机器学习算法用于模块化多电平换流器(modular multi-level converter，MMC)子模块的(sub-module，SM)开路故障诊断，并提取正常状态和故障模式下的三相交流电流和内环电流的样本数据来训练 MKSTM，从而实现开路故障的定位与诊断。

此外基于统计分析法、梯度下降学习、图学习、模糊逻辑、半监督学习、时间序列预测法等人工智能算法在电力电子变换器的故障诊断领域也得到广泛应用[71-76]。

## 2 基于人工智能的开路故障诊断应用实例

随着智能电网建设的不断推进，基于数据驱动的人工智能技术在电力电子和电气工程领域得到广泛应用。本节将对一些基于人工智能技术的电力电子变换器开路故障诊断方法及应用实例进行论述。

### 2.1 基于 SVM 的开路故障诊断方法

SVM 是一种建立在统计学习理论和结构风险最小化基础上的机器学习算法，可用于数据分析和模式识别，适用于小样本、非线性及高维度的模式识别、分类和回归分析[77-78]。其基本分类方法是采用非线性变换将一个输入空间变换到另一个高维空间，然后在这个高维的空间求出最优的线性分类界面。SVM 方法可以产生较为复杂的分界面，在特征多、类别结构复杂时仍有较高的分类精度。基于 SVM 的电力电子变换器故障诊断方法通常先将与故障相关的样本信号进行一系列特征变换或者特征提取，从而得到用于训练 SVM 故障诊断分类器的数据样本，再通过训练得到 SVM 故障诊断分类器来实现 IGBT 开路故障诊断及定位[79][80]。

文献[21]中针对风力发电系统中三相 PWM 变流器中 IGBT 开路问题进行研究(图 6)，以整流状态为例，选取直流侧输出电压信号进行分析，并利用小波包分析法对故障信号进行预处理，提取能量集中频带范围内的谱值作为故障特征数据样本，将该样本用于训练 SVM 故障诊断分类器，从而实现 PWM 整流器的故障诊断。根据文中研究可知，当不同 IGBT 发生开路故障时,输出电压差异并不大，而且大部分故障电压波形十分相似(图 7)，因此直接利用直流侧输出电压非常困难。

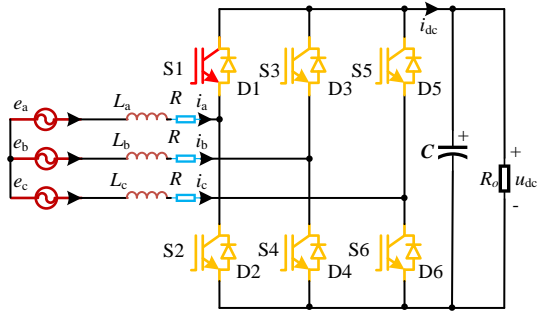

图 6 三相 PWM 整流电路

Fig. 6 Three-phase PWM rectifier circuit

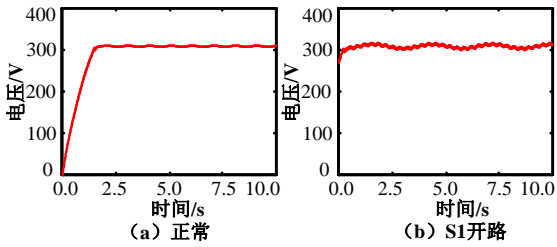

图 7 三相 PWM 整流器直流输出电压

Fig. 7 DC output voltage of three phase PWM rectifier

小波变换的滤波器组算法实现小波分解：对于任意长度为 $2^n$ 的输入序列 $a^n = \{a_{n,0},...,a_{n,2^n-1}\}$，利用求平均值和细节系数的方法，即对应低频分量 $a^{n-1}$ 和高频分量 $d^{n-1}$(从 0 开始计数)，其表达式分别为

$$\begin{cases} a^{n-1} = \{a_{n-1,0}, a_{n-1,1}..., a_{n-1,2^{n-1}-1}\} \\ d^{n-1} = \{d_{n-1,0}, d_{n-1,1}..., d_{n-1,2^{n-1}-1}\} \end{cases} \quad (1)$$

其中低频系数 $a_{n-1,k}$ 和高频系数 $d_{n-1,k}$ 分别为

$$\begin{cases} a_{n-1,k} = (a_{n,2k} + a_{n,2k+1})/\sqrt{2}, k=0,1,...,2^{n-1}-1 \\ d_{n-1,k} = (a_{n,2k} - a_{n,2k+1})/\sqrt{2}, k=0,1,...,2^{n-1}-1 \end{cases} \quad (2)$$

利用 db10 小波对三相 PWM 整流器直流侧输出电压信号进行 4 层小波包分解和重构后，正常模式和 S1 故障模式下的第 4 层第 1 个细节信号的频谱如图 8 所示，不同故障模式下不同频率的功率谱是不同的。通过小波包分析选择特征频率作为特征向量用于训练 SVM 分类器，从而实现故障诊断与定位(图 9)。

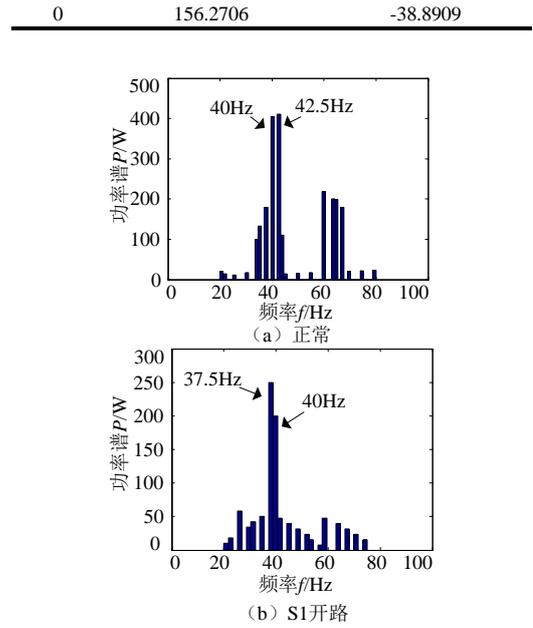

图 8 小波包重构后的第 4 层第 1 个细节信号的频谱

Fig. 8 The fourth floor first detail signals power spectrum of fault signals reconstructed by wavelet packet

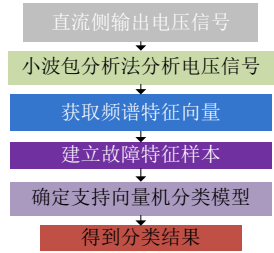

图 9 基于小波包分析与 SVM 的开路故障诊断流程图

Fig. 9 Flow chart of open-circuit fault diagnosis based on wavelet packet analysis and SVM

文献[79]通过分析发现子模块发生开路故障后交流侧的三相电流会产生相应的直流分量，并求取其包络均值作为最小二乘互信息谱聚类的训练集，将获得的聚类标签作为整体最小二乘支持向量机(total least square support vector machines，TLS-SVM)的训练样本标签，从而实现 MMC 变换器中子模块桥臂故障的定位(图 10、11)。由于高电平子模块数较多，未能准确定位出子模块故障位置。

表 1 一维哈尔小波变换的算例

Tab.1 Example of the one dimensional Haar wavelet transform

| 尺度 | 平均值 | 细节系数 |
|---|---|---|
| 3 | 48，34，24，60 | |
|  | 72，28，55，121 | |
| 2 | 57.982 8，59.397 0， | 9.899 5，-25.455 8， |
|  | 70.003 6，125.157 9 | 31.112 7，-46.669 0 |
| 1 | 83，138 | -1，-38 |
|  | | 0　　156.2706　　-38.8909 |

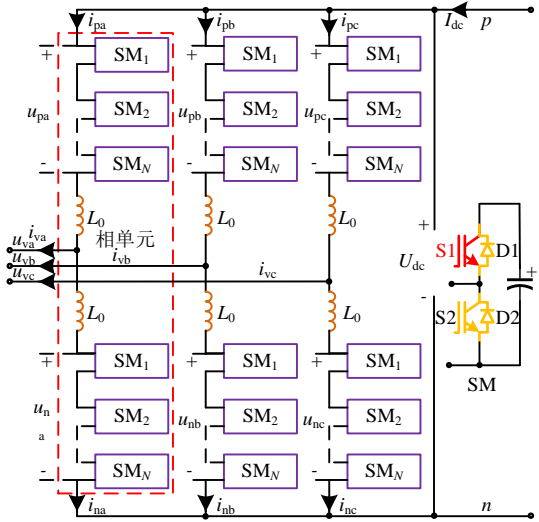

**图 10 MMC 原理图**

**Fig. 10 Diagram of MMC**

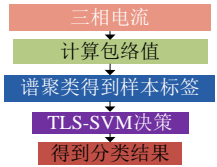

**图 11 MMC 子模块开路故障诊断流程**

**Fig. 11 MMC sub-module open-circuit faults diagnosis**

### 2.2 基于 ANN 的开路故障诊断方法

神经网络是一种模仿动物神经网络行为特征，进而建立分布式信息数据的广义数学模型，是联接主义智能实现的典范[81]。电力电子变换器故障诊断通常是利用神经网络算法对正常和故障状态的数据进行离线学习训练，通过调整神经元节点间的相互关系，通过网络层间的学习来建立故障特征与故障位置的映射关系，最终实现由故障特征到故障位置的推理过程，实现故障诊断和监测[82-85]。

文献[86]中大功率风力发电机组中 3 个并联电压源变换器的 IGBT 开路故障进行研究，电路基本结构如图 12 所示，根据开路故障短时间内危害较小的特点，采集一个周期的相电流作为样本特征来进行故障诊断，将三相电流进行 d-q 变换得到不同故障位置的故障电流矢量图，提取电流矢量表面积、矢量角、分布角(图 13)作为训练样本数据，然后利用神经网络进行学习训练实现变换器中 IGBT 开路故障的定位。

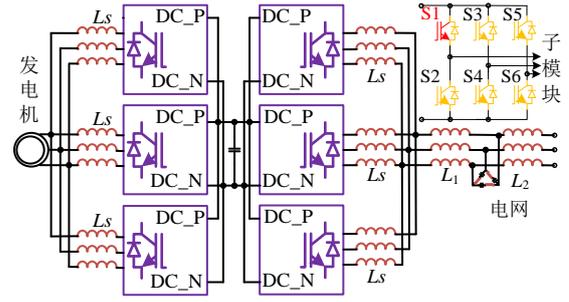

**图 12 三并联电压源变换器**

**Fig. 12 Three-Parallel voltage source converter**

首先将三相电流($i_a$，$i_b$，$i_c$)进行 d-q 变换，得到

$$\begin{bmatrix} i_d \\ i_q \end{bmatrix} = \begin{bmatrix} \frac{2}{3} & -\frac{1}{3} & -\frac{1}{3} \\ 0 & \frac{1}{\sqrt{3}} & -\frac{1}{\sqrt{3}} \end{bmatrix} \begin{bmatrix} i_a \\ i_b \\ i_c \end{bmatrix} \quad (3)$$

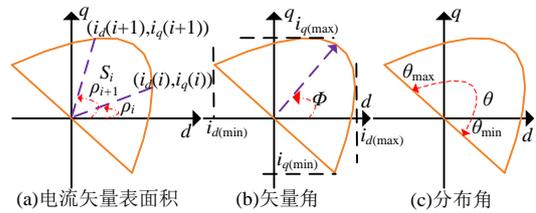

**图 13 提取的不同参数**

**Fig. 13 Different extracted parameters**

然后再得到 d-q 坐标系下不同故障模式下的电流矢量轨迹(图 14)。

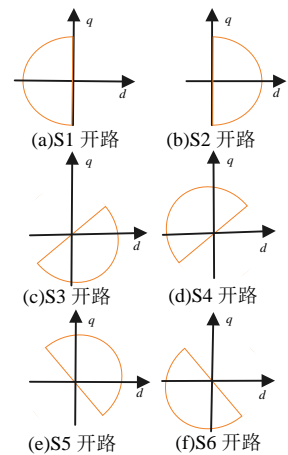

**图 14 不同故障模式下的电流矢量轨迹**

**Fig. 14 Current vector under different faults modes**

由图 14 可知，不同故障模式下具有不同的电流矢量轨迹，扇形的大小也就不相同，同时电流矢量表面积、矢量角、分布角也就不相同。

电流矢量表面积 $S$ 的表达式为

$$S = \sum_{i=0}^{n} S_i = \sum_{i=0}^{n} \frac{\pi r_i^2 \rho}{360} \quad (4)$$

式中：$r_i$ 为半径，$r_i = \sqrt{(i_{di})^2 + (i_{qi})^2}$，$i$ 为序号，$n$ 为总个数，d、q 为坐标轴，$\rho$ 为圆心角，$\rho = \rho_{i+1} - \rho_i$ (图 13(a))。

矢量角 $\phi$ 表达式为

$$\phi = \tan^{-1}\left(\frac{i_q}{i_d}\right) \quad (5)$$

分布角 $\theta$ 表达式为

$$\theta = \theta_{max} - \theta_{min} \quad (6)$$

$\theta_{max}$、$\theta_{min}$ 分别为一个周期内测量的分布角的最大值和最小值。

由于电流矢量表面积受到电流幅值变化的影响，单位电流矢量被用于解决该问题，单位电流矢量 $i_d'$ 与 $i_q'$ 表达式为

$$\begin{cases} i_d' = \dfrac{i_d}{\sqrt{(i_d)^2 + (i_q)^2}} \\ i_q' = \dfrac{i_q}{\sqrt{(i_d)^2 + (i_q)^2}} \end{cases} \quad (7)$$

将标记样本标签的单位电流矢量表面积、矢量角、分布角作为特征向量用于训练神经网络得到成熟的故障诊断分类器，当向成熟的故障诊断分类器输入未标记的样本时，就能够实现对不同开关管故障的定位及诊断(图 15)。

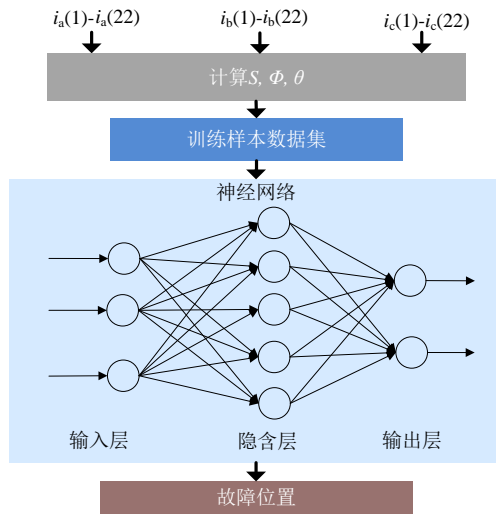

**图 15 基于神经网络的开路故障诊断方法**
**Fig. 15 Open-circuit fault diagnosis method based on Neural Network**

### 2.3 基于 RFs 的开路故障诊断方法

基于人工神经网络的故障诊断方法在学习训练过程中需要足够多的训练样本，而且容易陷入局部最优解；基于 SVM 的故障诊断方法在分类特征量较多的情况下，故障识别的速度变得较为缓慢，需要花费大量的时间、占用较多的资源，该算法不利于在线实现故障诊断；而随机森林算法可以用于处理多分类问题[23,87]。随机森林算法由 Breiman 在文献[88]提出，是基于多个随机决策树组成的 Bagging 类集成学习算法；其输出类别是根据所有决策树共同投票来决定，而且该算法具备参数量较少、不容易发生过拟合等优点[89-93]。

文献[25]针对双极性 SPWM 单相全桥逆变器中开关管的开路故障特征进行分析，并提取输出电压、输出电流和直流侧输入电流等作为故障诊断信号，然后计算它们的最大值、最小值、均值以及方差等作为特征样本，再利用随机森林来训练得到故障诊断分类器，从而实现逆变器的开路故障定位(图16)。

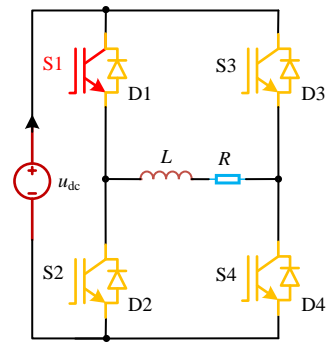

**图 16 双极性 SPWM 单相全桥逆变器**
**Fig. 16 Bipolar SPWM single-phase full-bridge inverter**

其提取特征的方式主要有以下 12 种方法：
1）最大值。

$$x_{max} = \max_{i \in \{1,2,\ldots,N\}}\{x_i\} \quad (8)$$

式中：$x_i$ 为一个周期中采集到的第 $i$ 个输出电压、输出电流或直流侧输入电流等诊断信号的数值；$N$ 表示一个周期内采集的样本数量。

2）最小值。

$$x_{min} = \min_{i \in \{1,2,\ldots,N\}}\{x_i\} \quad (9)$$

3）峰值。

$$x_{p-p} = x_{max} - x_{min} \quad (10)$$

4）平均值。

$$\bar{x} = \frac{1}{N}\sum_{i=1}^{N} x_i \quad (11)$$

5）方差。

$$\sigma^2 = \frac{1}{N}\sum_{i=1}^{N}(x_i - \bar{x})^2 \quad (12)$$

6）标准差。

$$\sigma = \sqrt{\frac{1}{N}\sum_{i=1}^{N}(x_i - \bar{x})^2} \quad (13)$$

7）峰度指数。

$$x_K = \frac{1}{N}\sum_{i=1}^{N}\left(\frac{x_i - \bar{x}}{\sqrt{\frac{1}{N}\sum_{i=1}^{N}x_i^2}}\right)^4 \quad (14)$$

8）偏度指数。

$$x_S = \frac{1}{N}\sum_{i=1}^{N}\left(\frac{x_i - \bar{x}}{\sqrt{\frac{1}{N}\sum_{i=1}^{N}x_i^2}}\right)^3 \quad (15)$$

9）波形指数。

$$K = \frac{\sqrt{\frac{1}{N}\sum_{i=1}^{N}x_i^2}}{\frac{1}{N}\sum_{i=1}^{N}|x_i|} \quad (16)$$

10）峰值指数。

$$C = \frac{x_{max}}{\sqrt{\frac{1}{N}\sum_{i=1}^{N}x_i^2}} \quad (17)$$

11）脉冲指数。

$$I = \frac{x_{max}}{\frac{1}{N}\sum_{i=1}^{N}|x_i|} \quad (18)$$

12）裕度指数。

$$C_{Lf} = \frac{x_{max}}{\left(\frac{1}{N}\sum_{i=1}^{N}\sqrt{|x_i|}\right)^2} \quad (19)$$

利用上述 12 种方法来提取输出电压、输出电流和输入电流的时域特征，3 种信号每周期各采集 200 个点，一组样本共有 36 种特征，按照图 17 所示来实现随机森林故障分类器的训练，从而实现开路故障位置的诊断。

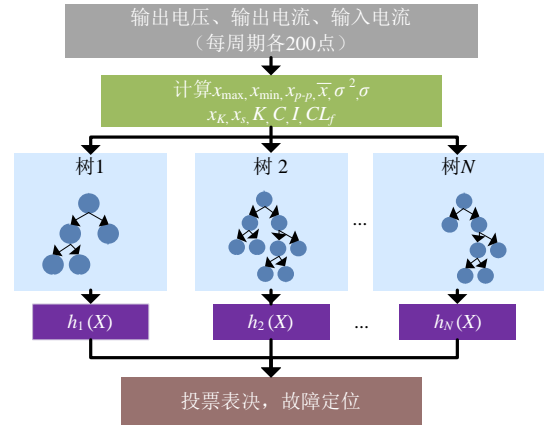

**图 17 基于随机森林的开路故障诊断方法**

**Fig. 17 Open-circuit fault diagnosis method based on random forests**

## 3 基于 RFs 与瞬时故障特征相结合的开路故障诊断方法

本文综述研究中，利用数据大多数为周期性的电流、电压等测量数据，而电力电子变换器故障时刻不同产生的周期性的故障样本也不相同，这样将会导致训练样本的数据量非常大。而采用瞬时故障样本就能够降低故障时刻的影响，同时也比较容易采集完备的瞬时故障特征样本。基于数据驱动的人工智能故障诊断方法仅需要利用故障状态下的历史数据进行训练就能实现对电力电子变换器的故障诊断和定位[94]。AC-DC 变换器被广泛用于直流输电、有源滤波器或电网接口中，本文提出基于随机森林与瞬时故障特征相结合的故障诊断方法，并以三相四线制 PWM 整流器和中点钳位式(neutral-point-clamped，NPC)逆变器为例进行验证。

为简化分析，三相四线制 PWM 整流器通常被看作 3 个独立的单相半桥整流器[95]。以 A 相为例进行分析，图 18 为 A 相器件在 $i_a$ 不同方向上的导通情况。如图 18(a)所示，当 $i_a \geq 0$ 时，S1 导通，S2 关断，$i_a$ 通过 D1 实现正向续流；S1 关断，S2 导通时，$i_a$ 通过 S2 实现正向续流。在上述两种情况下，S1 没有参与 $i_a$ 的流通，S1 故障没有影响，而 S2 就会有影响。如图 18(b)所示，当 $i_a < 0$ 时，S1 导通，S2 关断，$i_a$ 通过 S1 实现反向续流；S1 关断，S2 导通时，$i_a$ 通过 D2 实现反向续流。在上述 2 种情况下，S1 参与 $i_a$ 的流通，S1 故障会有影响，而 S2 故障没有影响。因此，当 S1 故障时，$i_a$ 的波形没有负半周，当 S2 故障时，$i_a$ 的波形没有正半周(图 19)。当开路故障时，所配备的二极管能够继续作为整流元件工作，输出电压会出现波动，但是系统不会立即崩溃。

可以看出，不同位置 IGBT 开路故障后，对应三相电流波形的变化也会不同，因此可以考虑利用三相电流作为故障诊断信号来实现 IGBT 开路故障的诊断及定位。

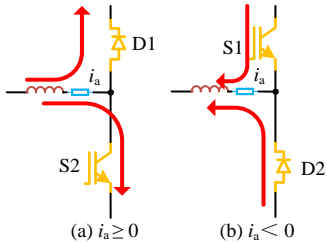

图 18 $i_a$ 不同方向的导通情况

Fig. 18 Conducting route under different directions of $i_a$

提取各种故障状态下的三相瞬时电流数据作为故障样本，按照表 2 所示对故障样本进行标记。在故障诊断分类器的训练过程中(图 20)，从不同故障状态组成的 24000 个样本中随机选取 8000 个样本构成训练集，其余样本作为测试样本，其瞬时诊断样本格式为 $(x(t), y(t))$，$x(t)$ 为 $t$ 时刻的故障电流样本($i_a(t)$，$i_b(t)$，$i_c(t)$)，$y(t)$ 为该时刻的样本类别标签，将故障电流进行归一化处理后与样本类别标签合并构成所有的样本训练集。采用 Bagging 类算法随机抽取样本，随机森林算法中引入随机特征来构成子样本训练集；然后采用 CART 树模型来训练基模型分类器，通过随机的形式训练得到不同的多棵 CART 决策树。然后按照多数投票法对所有决策树进行集成得到随机森林故障诊断分类器。故障诊断时只需要输入未知类别的三相瞬时故障电流样本数据，随机森林故障诊断分类器即可输出该样本的类别，从而实现开路故障位置的诊断。训练过程采用交叉验证法，分类正确率达到 98.72%。随机森林决策树的个数为 264 时取得最高分类精度(图 21)，部分样本及诊断结果如表 3 所列。

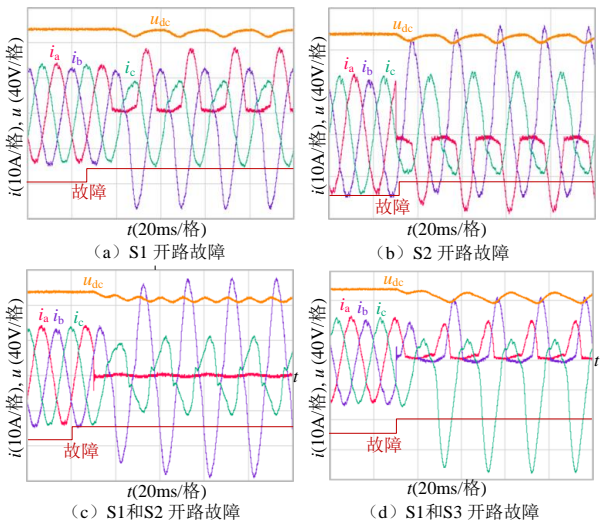

（a）S1 开路故障　　　　（b）S2 开路故障

（c）S1和S2 开路故障　　（d）S1和S3 开路故障

图 19 部分 IGBT 开路故障电压电流波形

Fig. 19 Voltage and current waveform when open-circuit faults occur in some IGBTs

表 2 故障 IGBT 和分类标签
Tab.2 Fault IGBT and classification labels

| 故障位置 | 分类标签 | | | | | |
| --- | --- | --- | --- | --- | --- | --- |
| | d1 | d2 | d3 | d4 | d5 | d6 |
| 正常 | 0 | 0 | 0 | 0 | 0 | 0 |
| S1 | 1 | 0 | 0 | 0 | 0 | 0 |
| S2 | 0 | 1 | 0 | 0 | 0 | 0 |
| S3 | 0 | 0 | 1 | 0 | 0 | 0 |
| S4 | 0 | 0 | 0 | 1 | 0 | 0 |
| S5 | 0 | 0 | 0 | 0 | 1 | 0 |
| S6 | 0 | 0 | 0 | 0 | 0 | 1 |
| … | | | … | | | |

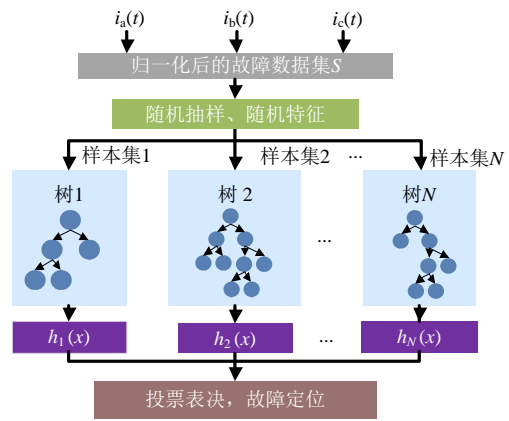

图 20 基于 RFs 与瞬时故障特征的开路故障诊断方法

Fig. 20 Open-circuit fault diagnosis method based on RFs and transient fault features

表 3 部分故障样本及诊断结果
Tab.3 Some fault samples and diagnosis results

| 故障位置 | $i_a$ | $i_b$ | $i_c$ | 实际类型 | 诊断类型 |
| --- | --- | --- | --- | --- | --- |
| 正常 | 0 | -14.280 | 13.615 | 000000 | 000000 |
| S1 | 0.332 | 12.287 | -12.619 | 100000 | 100000 |
| S1 | -0.332 | 7.638 | 2.324 | 100000 | 100000 |
| S2 | -2.324 | -13.6158 | 14.280 | 010000 | 010000 |
| S2 | -1.328 | -9.962 | -0.664 | 010000 | 010000 |
| S3 | -12.287 | 0.664 | 13.947 | 001000 | 001000 |
| S1 和 S3 | 0.664 | -0.332 | 5.645 | 101000 | 101000 |
| S1 和 S3 | 0 | 0.996 | 4.649 | 101000 | 101000 |
| … | … | | | | |

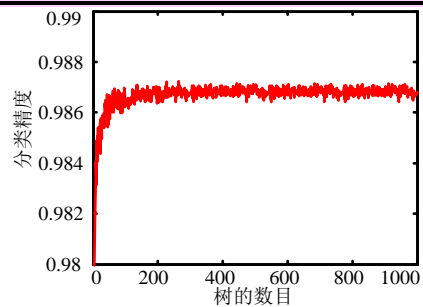

图 21 决策树数目对分类精度的影响

Fig. 21 Influence of decision tree on performance

由特征分析可知三相四线制 PWM 整流器中上

桥臂 IGBT 开路故障影响三相电流负半波的波形，下桥臂 IGBT 开路故障影响三相电流正半波的波形。根据以上规律可知理想状态下三相电流故障波形时间分布如图 22 所示，图中 SI 区域能够诊断出 S2，S3，S6 的任意组合，SII 区域能够诊断出 S2，S4，S6 的组合，同理 SIII，SIV，SV，SVI 区域也有对应组合。如 S1，S3 在 SI 区域同时发生故障时，S1 为 A 相上桥臂 IGBT，SI 区域 A 相电流为正半波，故障特征并不会表现出来，则利用瞬态故障特征就不能检测出 S1 也发生开路故障。因此本文提出一种基于多时间尺度的故障诊断方法来弥补多个开关管故障后而瞬时特征不同时出现的问题。

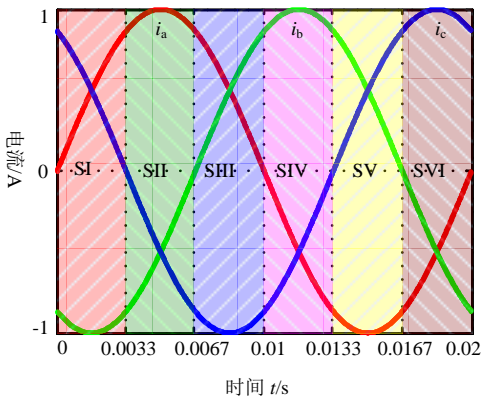

**图 22 故障特征分布**
**Fig. 22 Distribution of faults features**

图 23 为三相四线制 PWM 整流器开路故障诊断系统框图，图 24 为故障诊断实验过程，图 23 中闭环控制系统的时钟 clk1 为 128 kHz，采样频率 $f_{sampling}$ 为 25.6 kHz；然后再以时钟 clk2 为 10 kHz，重采样频率 $f_{resampling}$ 为 10 kHz 对三相电流信号进行重采样。因此，它每周期(20 ms)只向工控系统发送 200 组故障样本，该方法能够降低数据传输的压力。采用多时间尺度的故障诊断方法既解决了部分 IGBT 故障后瞬时故障特征不同时出现的问题，同时也解决了大多数人工智能算法更适合在计算机上运行的问题。基于随机森林算法与瞬时故障特征训练得到的成熟的故障诊断分类器每周期可输出 200 个诊断结果，当 S1 和 S3 同时发生故障后的诊断结果如图 25 所示，200 个诊断结果共同决定故障位置，图 25 中诊断结果的顺序为 S3 开路故障→正常状态→S1 开路故障→S1 和 S3 开路故障→S3 开路故障，在诊断过程中出现了部分样本被诊断为 S4 和 S5，但是该结果仅出现了很短时间而且后续不再出现，因此，最终诊断结果表明 S1 和 S3 都发生开路故障。一旦检测到开路故障，在线故障诊断系统将向 FPGA(field-programmable gate array)控制器发送保护信号，从而为后续的容错处理或者切换备用设备奠定基础。

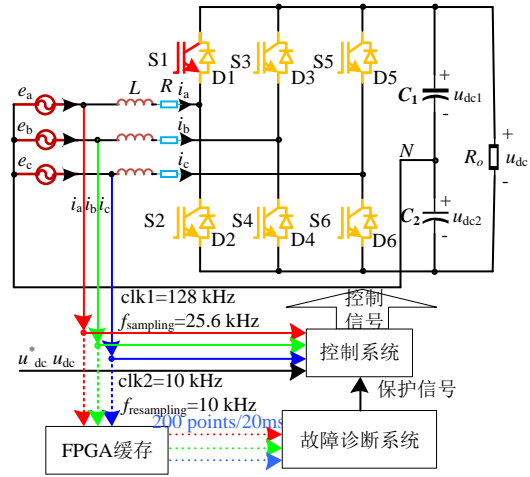

**图 23 三相四线制 PWM 整流器故障诊断系统**
**Fig. 23 Diagram of fault diagnosis system for three-phase four-wire PWM rectifier**

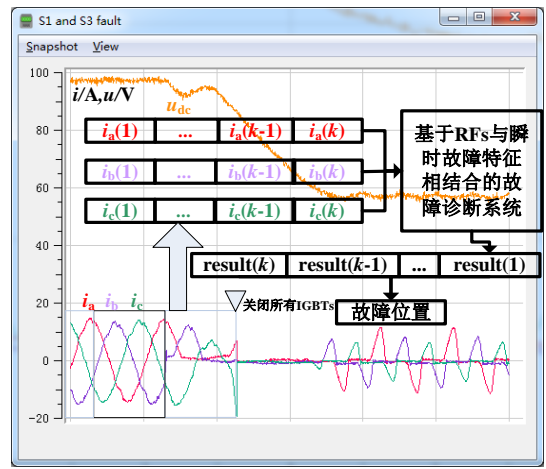

**图 24 故障诊断实验**
**Fig. 24 Fault diagnosis Experiments**

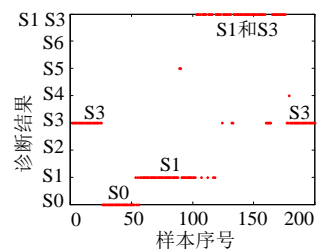

**图 25 故障诊断结果**
**Fig. 25 Faults diagnosis results**

此外该方法在 NPC 逆变器的 IGBT 开路故障诊断中也进行验证，NPC 逆变器电路结构如图 26 所示，其中 A 相 IGBT 开路故障后电流波形如图 27 所示。由于该方法使用的是三相电流为故障样本，因此该方法也适用于其它多数三相电力电子变换系

统的开路故障诊断。

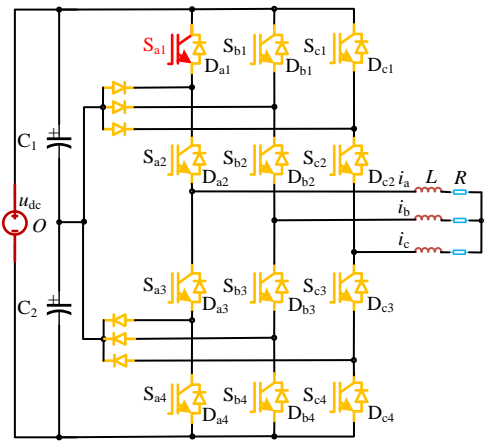

**图 26 NPC 逆变器**
**Fig. 26 NPC inverter**

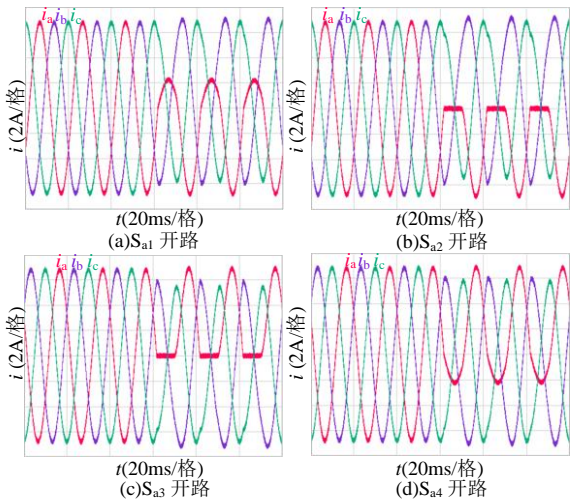

**图 27 A 相 IGBT 故障电流**
**Fig.　27 A-phase IGBT fault current waveform**

## 4 结论

本文针对电力电子变换器故障特征进行分析，并对基于人工智能技术的电力电子变换器开路故障诊断方法进行综述，并给出具体的应用实例。然后提出一种基于随机森林与瞬时故障特征相结合的开路故障诊断方法，并应用于三相电力电子变换器的开路故障诊断，对提高电力电子系统的稳定性具有重要意义。本研究针对人工智能技术在电力电子变换器开路故障诊断领域面临的挑战，提出以下研究要点及趋势：

1）随着电力电子变换器在新能源领域的广泛应用，测量设备逐渐增加，采集样本的数量和维度越来越大，对于数据的压缩、传输、存储带来困难。考虑针对不同电力电子变换器的故障特征分析，利用相关性分析方法、小波压缩等方法去除冗余特征、冗余数据等，从而实现数据维度和数据量的压缩。

2）由于电力电子变换器运行环境复杂，负载波动变化等影响，使得采集完备的故障样本难度加大。可以考虑基于数据驱动与知识驱动相结合的故障诊断方法，利用电力电子变换系统中的现有知识对采集样本进行特征变换，提取出不受外界干扰的样本特征，降低对历史数据样本的过度依赖；同时还可以考虑将各种信号处理方法应用于多源、多时间尺度信号的特征提取中，实现多信息融合的电力电子变换器故障诊断，充分发挥不同故障诊断方法的优势。

3）针对基于解析模型的电力电子变换器故障诊断方法中参数设定复杂的问题，可以考虑利用人工智能算法来实现参数的设定，降低参数设定的难度，使得参数更准确可靠。